\documentclass[11pt,twoside]{article}

\usepackage{asp2006}
\usepackage{epsf}
\usepackage{psfig}
\usepackage{lscape}
\usepackage{graphicx}
\begin{document}
\title{The Magellanic Stream to Halo Interface:
Processes that shape our nearest gaseous Halo Stream}   

\author{
Lou Nigra,\altaffilmark{1}$^{,}$\altaffilmark{2}
Sne\v{z}ana Stanimirovi\'{c},\altaffilmark{1} 
J. S. Gallagher, III,\altaffilmark{1}
\\Felix J. Lockman,\altaffilmark{3}
David L. Nidever,\altaffilmark{4}
Steven R. Majewski\altaffilmark{4}
} 

\altaffiltext{1}{Department of Astronomy, University of Wisconsin, Madison WI 53706, USA}
\altaffiltext{2}{National Astronomy and Ionosphere Center, HC3 Box 53995, Arecibo, PR 00612 USA}
\altaffiltext{3}{National Radio Astronomy Observatory, Green Bank, WV 24944-0002, USA}
\altaffiltext{4}{Department of Astronomy, University of Virginia, Charlottesville, VA 22904-4325, USA}

\begin{abstract} 
Understanding the hydrodynamical processes and conditions at the 
interface between the Magellanic 
Stream (MS) and the Galactic halo is critical to understanding
the MS and by extension, gaseous tails in other interacting galaxies. 
These processes operate on relatively small scales and not only help shape 
this clumpy stream, but also affect the neutral gas dynamics and transfer of 
mass from the stream to the halo, thus affecting 
metal enrichment and gas replenishment of the Galaxy.
We describe an observational program to place constraints on these processes 
through high-resolution 
measurements of HI emission, HI absorption 
and H$\alpha$ emission with unprecedented sensitivity.
Methods will include structural analysis, searching for cold gas cores in 
clumps  and analyzing gas 
kinematics as it  transitions to the halo. The latter method includes sophisticated 
spatial integration techniques to 
deeply probe the neutral gas, which we apply to a new HI map obtained from 
the Green Bank Telescope  
with the highest sensitivity HI observations of the MS to date. 
We demonstrate that the integration techniques enhance 
sensitivity even further, thus allowing detection of apparent MS gas components with 
density approaching that of the Galactic halo.
\end{abstract}

\section{Introduction}  
The Magellanic Stream (MS) is our nearest major gaseous interaction remnant. 
It trails across much of the 
southern galactic sky for $>100^{\circ}$ behind the Magellanic Clouds (MCs), passing near the 
Southern Galactic Pole at its midpoint. The MS head is presumed to be nominally between $52$ 
and $61$~kpc away, the distances to the MCs established by \citet{koerwer09} and \citet{hilditch05}. The
distance to the tip
is not well-constrained but estimates are of the same order as distances to the MCs. 
These distances are about 15 times closer than the nearest extragalactic tidal features of the 
Andromeda system, therefore the MS presents us with the opportunity to closely study star formation 
potential (or lack thereof), mass transfer, and kinematics of a 
major interaction feature on fine scales 
inaccessible elsewhere.
No signs of stars or even star formation have been found in the MS to date. 
The mean column density of the MS ranges from $4\times10^{18}$ to $4\times10^{19}$~cm$^{-2}$ 
along its length, which is an order of magnitude 
below the tidal debris star-formation threshold ($4\times10^{20}$ cm$^{-2}$) 
found by \citet{maybhate07}. Unlike 
High Velocity Clouds (HVCs), whose H$\alpha$ emission can be explained by 
Galactic UV radiation, the MS is brighter in H$\alpha$, particularly near the MCs, suggesting 
another ionizing source. Global 
models of the MS using Smoothed Particle Hydrodynamics (SPH) and N-Body techniques have
reproduced its large-scale 
structure with some degree of success with the then-current assumption of multiple MC orbits of
the Galaxy 
\citep{moore94, connors06, mastro05}. The new measurement of MC proper motions by \citet{kalliv06a} 
and \citet{kalliv06b} led \citet{besla07} to propose the likelihood of unbound MC orbits. 
Subsequently, \citet{mastro08} presented revised results for the new orbits.

Something that the global models do not capture is the very rich, 
fine-scale structure revealed in recent 
HI MS maps such as  \citet{putman03b}  and \citet{bruns05}. Even finer 
structure was revealed by \citet{stanim08} in a high-resolution map of 
the northern tip of the MS using the 305$\:$m dish at Arecibo. 
These observations revealed extended fine filamentary structure and clumps down to 
the angular resolution of the telescope ($3.5'$). 
They also highlighted the importance of  
Kelvin-Helmholtz and Thermal Instabilities in 
forming the clumpy structure of the MS. 
In addition, they revealed cooler cores in some clumps 
suggesting star formation potential and providing clues to halo pressure 
and dark matter confinement of these clumps.
Processes are clearly at work on these small scales that could affect star formation, the transfer of gas to 
the halo, and also may provide additional drag affecting MS global dynamics.

The global simulations 
mentioned above can miss these processes since SPH can suppress instabilities due to smoothing 
\citep{agertz07}, and N-body simulations ignore gas processes altogether. Grid-based modeling can 
capture these small-scale instabilities better, exemplified by simulations 
exploring mechanisms for 
excess H$\alpha$ emission in the MS \citep{jbh07}, galaxy replenishment
 \citep{jbh09,heitsch09}, and 
HVCs in the halo \citep{quilis01}. Grid-based modeling on these scales is producing some exciting 
insights into the possible processes at work. 

We present here a research program aimed at setting some 
observational constraints on these processes that act on the MS at its interface with the Galactic halo and 
then using them to test these models.
Towards this end, we have obtained the most sensitive HI emission map
of portions of the MS to date and have begun to analyze the structure and kinematics of the gas.
We are in the process of obtaining 
high resolution, high sensitivity measurements of H$\alpha$ emission for insight into higher energy
processes at the halo interface, and HI absorption to look for cold gas 
cores and characterize gas temperatures.

\section{Observations: Completed, ongoing and planned}
We have completed a program at the Green Bank Telescope
\footnote{
The Robert C. Byrd Green Bank Telescope is operated by the
National Radio Astronomy Observatory, which is a facility of
the US National Science Foundation operated under cooperative
agreement by Associated Universities, Inc.
}
(beamwidth~$=9.1'$)
where we obtained deep HI spectra across two separate $12$~deg$^2$ regions. One region (Region 1) 
is located in the northern tip and the other (Region 2) is $~20$~deg further up in the mid-MS region. 
On-the-fly (OTF) mapping and in-band frequency switching was employed.
Region 1 data have been reduced with a fairly successful initial pass at baseline removal resulting
in a cube with $3.53'$ square pixels, and
velocity resolution of $0.161$~km$\:$sec$^{-1}$ from $-200$ to $-518$~km$\:$sec$^{-1}$.
Noise was measured at 
$\sigma_T=4.2$~mK corresponding to column 
density noise $\sigma_N=1.14\times10^{17}$~cm$^{-2}$ for a $15$~km$\:$sec$^{-1}$ FWHM profile. 
At less than half the noise of the Galactic
All-Sky Survey (GASS) \citep{mg09} scaled to the same profile, the map presented here is
the most sensitive of the MS to date.
Column density and velocity field maps are shown in Figure~\ref{fig:region1map}.

Further plans include observations of the 
fine-scale signature of ionization process on selected clumps and features by obtaining deeply integrated,
high-resolution H$\alpha$ observations.  We will compare our results to those expected in models of 
ionization through Galactic UV radiation or energetic gas processes such as those proposed by 
\citet{jbh07}.
We also plan to obtain deep absorption measurements in the direction of 
several background radio sources within the tip of the MS 
coinciding with clumps identified by \citet{stanim08}.
In addition to possibly identifying cold, potentially star-forming cores, 
these sensitive absorption measurements through specific small-scale structures 
will provide useful constraints on the analysis of gas kinematics.

\section{Results \& Discussion}
\begin{figure}
\begin{center}
\includegraphics[height=3.4in]{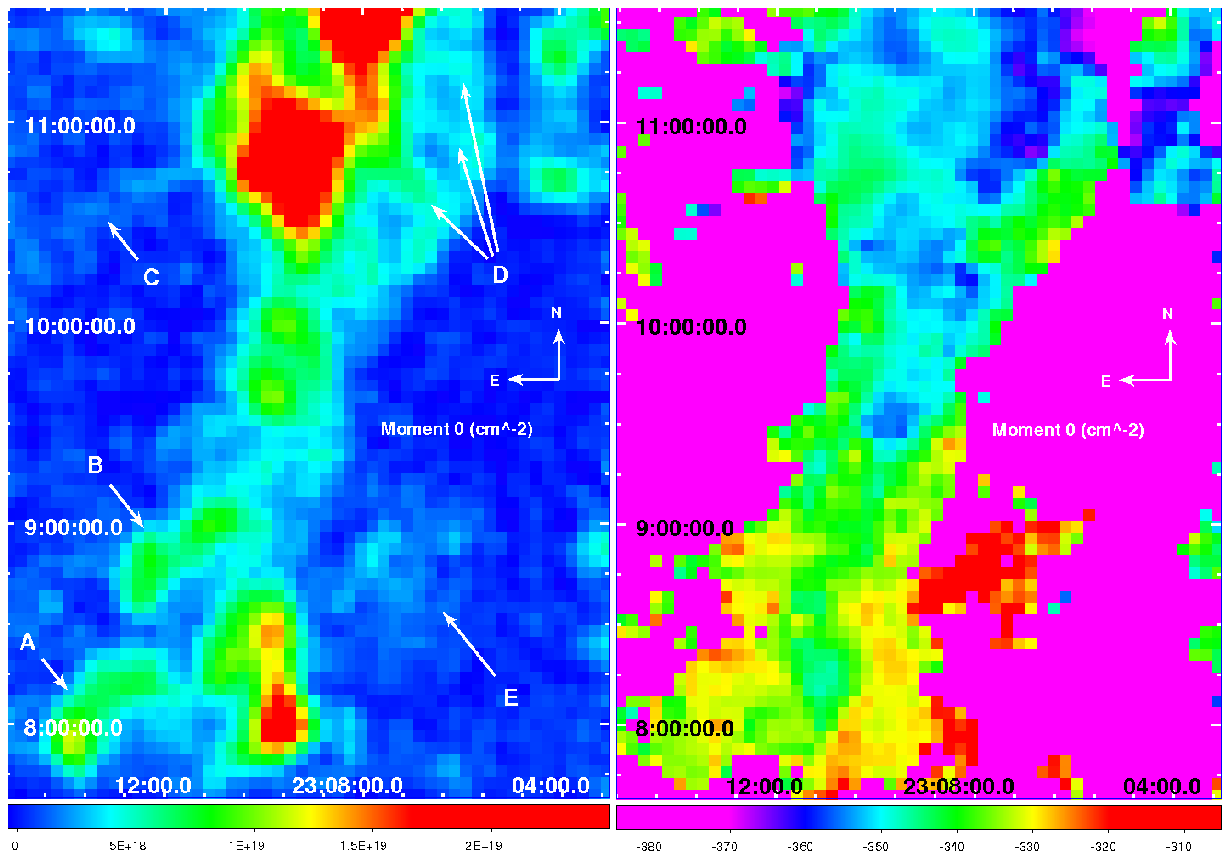}
\caption{From \citet{nigra09}. Left - Region 1 column density map from the
GBT data. Velocity range is $-385$ to $-305$~km$\:$sec$^{-1}$.
Intensity scale is $0\:$to$\:2.5\times10^{19}\:$cm$^{-2}$. 
Features ``A'' through 
``E'' are discussed in the text. Right - Region 1 velocity field for 
$N_{HI}>1.5\times10^{18}$~cm$^{-2}$. 
See text for discussion. These images were generated with data from 
telescopes of the National Radio 
Astronomy Observatory, a National Science Foundation Facility, 
managed by Associated Universities. Inc.}
\label{fig:region1map}
\end{center}
\end{figure}

The Region 1 HI column density image 
of Figure~\ref{fig:region1map} (left), shows
the main filament extending from the lower left to the upper right, which
is part of the longer filament S2 identified by \citet{stanim08}.  
The Region 1 velocity field (Figure~\ref{fig:region1map} right)
shows the general velocity gradient of the main filament, but also 
reveals some differentiation in places, 
suggesting some complexity in the projected dimension.
Several prominent and interesting features are labeled in Figure 1:
(A) and (B) appear as head-tail clumps;
(C) is a narrow filament that apparently connects to the main filament;
(D) are a series of relatively dense ``spokes'' at the edge of the main filament;
(E) is a large, apparently coherent cloud that is significantly 
more diffuse than other clumps of its size in the map. 
Most of these features (A through D) have a strong component transverse to the main 
filament, suggesting that this filamentary substructure 
may represent ram-pressure or ablation shreds coming off the main
cloud. It is interesting to note that most of the transverse 
substructure appears in the form of short filaments, instead of diffuse cloud envelopes. 
In the future, these structures will be analyzed further both spatially and 
kinematically to see what the de-projected structures 
might look like and what they might say about the 
processes at work here. We will 
also statistically compare the structure of this region  to 
simulations from \citet{jbh07} by using Fourier 
Transform methods.

\begin{figure}
\begin{center}
\includegraphics[height=2.4in]{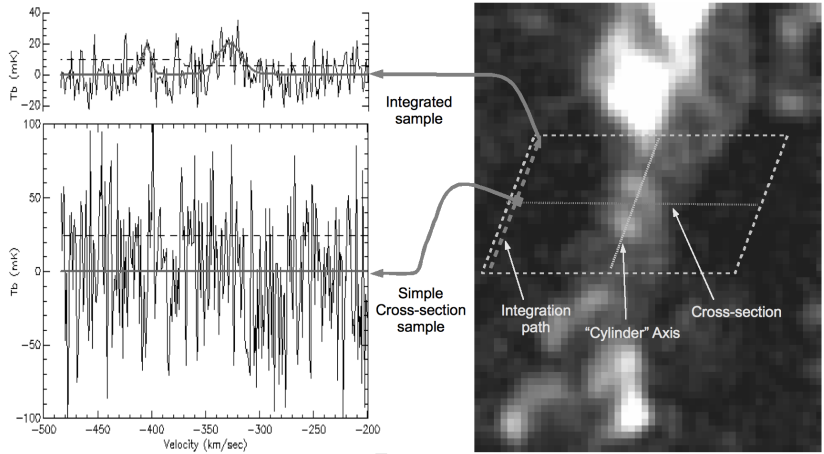}
\caption{ A demonstration of integration techniques used for gas analysis. 
A portion of the main 
filament is modeled as a cylinder. Integrating (averaging) parallel to 
its axis some distance away 
has two clearly detectable components while a single pixel spectrum at 
that distance has none.}
\label{fig:integrate}
\end{center}
\end{figure}

We are in the process of characterizing
properties of the diffuse neutral gas in the MS-halo transition region.  
Profiles of gas properties (temperature and column density) 
as a function of distance from the clump 
centers will be compared with analytical and simulation results for various 
hydrodynamic processes, each of which will have a 
particular ``signature'' profile. 
For instance, the temperature vs. radial distance in an isolated cloud in the 
halo is quite different if experiencing  saturated 
evaporation or if radiatively stabilized, as shown by \citet{cowie77} and \citet{mckee77}. 

These profiles are conventionally obtained by looking at observed spectra along a 
cross-section and fitting to a curve, as in \citet{bruns01}. 
Further from the center, brightness decreases and velocity dispersion 
increases, causing measurement uncertainty to increase rapidly. 
In order to improve upon this method, we average 
spatially along a symmetry dimension under the assumption that the clump 
symmetry extends past its apparent boundaries. 
For instance, an isolated circular clump in a map is 
assumed spherical and can be integrated along concentric rings where 
each column of gas is nominally the same. Similarly for a 
linear feature, we assume cylindrical symmetry and 
average along lines parallel to the filament. 
More advanced techniques include integrating along $N_{HI}$ contours or 
optimizing to fit a parametric 3D gas model.

To demonstrate the sensitivity of this approach, we show initial results from 
applying this method in Figure~\ref{fig:integrate}. 
Here a portion of the main filament of Region 1 is assumed to be roughly 
cylindrical along the axis indicated on the image. 
The spectra along a parallel path to the left were 
integrated (averaged) after removing the velocity gradient of the main filament. 
The spectrum sampled at a a single location, as used in a simple cross-section profile, 
shows no sign of a detectable gas component, while the corresponding
averaged spectrum
of the cylindrical model clearly shows two gaussian components, with 
FWHM~$=10~$and$~26$~km$\:$sec$^{-1}$, respectively. Each is
detected well above their $3\sigma$ detection limits (dashed line) 
with corresponding $N_{HI}=3.4\times10^{17}$~cm$^{-2}~$and$~1.1\times10^{18}$~cm$^{-2}$,
respectively. 

Although detection of these apparent gas components
clearly demonstrates
the sensitivity of the technique,
verification as actual low column density gas is still in progress.
Their shape, width and
velocity are roughly consistent with MS
gas, and are not likely artifacts of the baseline
procedure. However, at  such high sensitivity, systematic artifacts not
normally seen must be considered \citep{nidever09}. 
Should either of these detections prove valid,
it would suggest that a significant amount of low column density
HI gas, well below thresholds of previous observations,
exists in this part of the MS. If not, it will still establish unprecedented
limits on a diffuse neutral gas component.
In either case, this approach will provide valuable insight into the history, dynamics
and hydrodynamic processes  of the MS \citep{nigra09}.

To gauge how close these detections are to the halo interface, an order 
of magnitude estimate of the number density is made. 
The filament's scale is $\sim1^{\circ}$ 
on the sky. Assuming a distance of $\sim60$~kpc, the scale width is then
$\sim1$~kpc$\:=3.1\times10^{21}$~cm. The 
number density for the $26$~km$\:$sec$^{-1}$ FWHM component is then on the order of 
$1.1\times10^{18}\div3.1\times10^{21}=3.5\times10^{-4}$~cm$^{-3}$. 
This is on the same order as the upper range of ionized halo number density 
estimates \citep{sembach03} and we can detect even weaker lines, so we may 
indeed be probing HI very close to the halo interface.

\section{Conclusions}
The rich fine, clumpy and filamentary structure on arcminute scales revealed in modern HI maps of the 
MS are not reproduced in global SPH and N-body simulations. Observations and grid-based local 
simulations reveal that this structure is likely produced by hydrodynamical instabilities at the interface 
between the cool, stripped gas as it moves through the hot halo gas. These processes may be an 
important source of MS structure, a mechanism for ionizing the MS gas as well as a significant means of 
transferring gas from the MS to the halo, and eventually to the Galactic disk.

A program is in progress to place observational constraints on these processes and on the modeling of them
by obtaining the most sensitive measurements to date of the MS gas in both HI and H$\alpha$ 
emission, as well as in HI absorption, on the periphery of the MS where it is closest to the ambient halo 
environment.  We have already obtained maps of two regions of the MS with unprecedented sensitivity. 
Using these data, we have demonstrated spatial integration techniques allowing detection of HI 
components at densities of $\sim10^{-4}$~cm$^{-3}$, within an order of magnitude 
of estimated halo density upper limits. 
We will apply these techniques to characterize gas 
kinematics and then compare to models to determine what processes are operating on the MS 
periphery.

\acknowledgements The authors thank Carl Heiles for his invaluable participation in the absorption
measurement program.
LN thanks the National Astronomy and Ionosphere Center for pre-doctoral support.
JSG thanks the University of Wisconsin Graduate School for partial support of this research.

\end{document}